\documentclass[twocolumn,superscriptaddress,floatfix,preprintnumbers,amssymb ,amsmath]{revtex4-1}
\usepackage{graphicx,amssymb}
\usepackage{color,ulem}\usepackage{bm}  
\usepackage{float}

\begin{document}
\title{Observing temperature fluctuations of a mesoscopic electron system}
\author{Bayan Karimi}
\affiliation{QTF Centre of Excellence, Department of Applied Physics, Aalto University School of Science, P.O. Box 13500, 00076 Aalto, Finland}
\author{Fredrik Brange}
\affiliation{Department of Physics and NanoLund, Lund University, Box 188, SE-221 00 Lund, Sweden}
\author{Peter Samuelsson}
\affiliation{Department of Physics and NanoLund, Lund University, Box 188, SE-221 00 Lund, Sweden}
\author{Jukka P. Pekola}
\affiliation{QTF Centre of Excellence, Department of Applied Physics, Aalto University School of Science, P.O. Box 13500, 00076 Aalto, Finland}

\date{\today}


\maketitle

\begin{figure*}
	\centering
	\includegraphics [width= \textwidth] {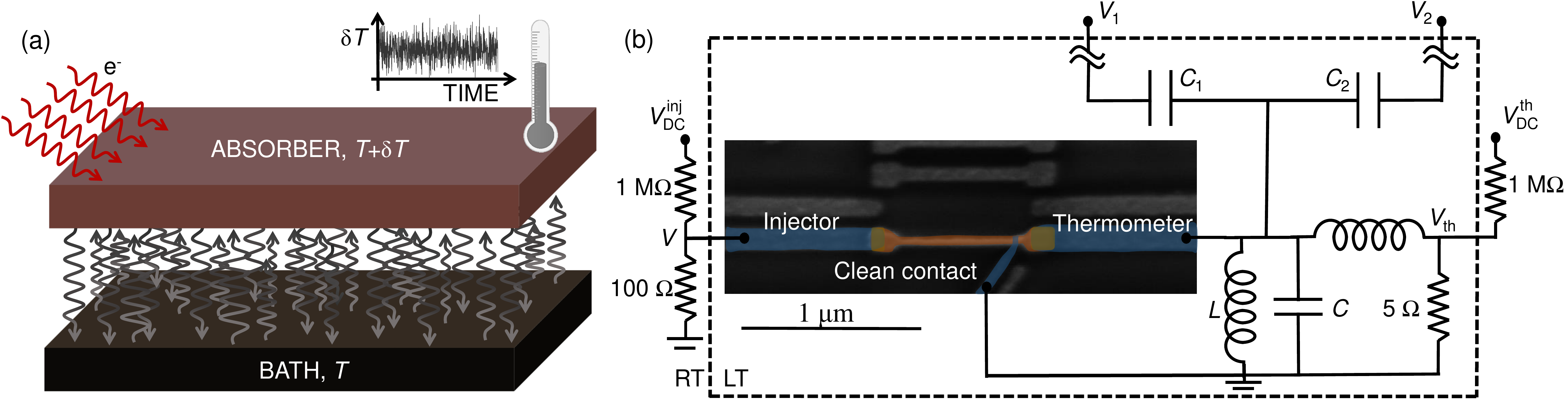}
	\caption{The setup for measuring temperature fluctuations. (a) The calorimeter principle applied to the electronic system in this work. The normal-metal absorber in the center is subjected to the fluctuating heat current from the phonon bath below. Additionally we have an option to create nonequilibrium by injecting "hot" electrons as indicated by red arrows on the left. A key element in the calorimeter is a thermometer with sufficient bandwidth to provide temporal temperature traces, of which an example is shown above the absorber. (b) The measurement setup including the colored scanning electron micrograph of the sample in the centre. The $L=1~\mu$m long Cu absorber (brown) coupled to two superconducting Al leads (blue) via tunnel barriers (bronze). The clean metal-to-metal contact to another superconducting Al lead pointing down at an inclined angle provides the proximity effect for the thermometer and a fixed chemical potential for the absorber. The circuit on the sample stage at low temperature (LT) within the dashed area presents the RF readout of the thermometer junction composed of an LC resonator and probed by RF transmission measurement between ports $V_1$ and $V_2$. The rest of the setup at room temperature (RT) is for DC biasing of both the  injector $(V)$ and thermometer $(V_{\rm th})$.}
	\label{fig1}
\end{figure*}
Almost a century ago, Johnson and Nyquist~\cite{johnson,nyquist} presented evidence of fluctuating electrical current and the governing fluctuation dissipation theorem (FDT). Whether, likewise, {\it temperature} ${T}$ can fluctuate is a controversial topic and has led to scientific debates for several decades~\cite{lifshitz,kittel,mandelbrot}. To resolve this issue, there was an experiment initially in 1992~\cite{lipa,lipa2} where the authors found good agreement between the FDT theory for heat and experiment on a macroscopic sample. A key question is what happens when we consider a nanoscale system with much fewer particles at 100 times lower temperatures. This challenge has not been addressed up to now, due to the demanding experimental requirement on fast and sensitive thermometry on a mesoscopic absorber. Here we observe equilibrium fluctuations of temperature in a canonical system of about $10^8$ electrons exchanging energy with phonon bath at a fixed temperature. Moreover, temperature fluctuations under nonequilibrium conditions present a nontrivial dependence on the chemical potential bias of a hot electron source. These fundamental fluctuations of $T$ set the ultimate lower bound of the energy resolution of a calorimeter.

Consider fluctuations of a system with coupling to a heat bath at temperature $T$ for which the classical FDT of heat current $\dot Q$ holds in form $S_{\dot Q}^{\rm eq} = 2k_BT^2 G_{{\rm th}}$ in equilibrium. Here $G_{\rm th}$ is the heat conductance to the bath. We can write the energy balance equation $\dot Q = C d{\widetilde T}/dt$ for the temperature of the system ${\widetilde T}(t)=T+\delta T(t)$ at time $t$, where $C$ denotes the heat capacity. The heat current is composed of its expectation value $-G_{\rm th}\delta T$ and fluctuations $\delta \dot Q$ around it. There are two origins of noise in this heat current: 1. there is standard randomness of a transport known for particle current noise (time randomness), and 2. the energies exchanged are also to some extent random leading to enhancement of fluctuations on top of those known for particle current only. We obtain the noise spectrum of temperature of the system by Fourier transformation as $S_T(\omega) =\int dt e^{i\omega t}\langle \delta T(t)\delta T(0)\rangle$. This yields under steady state conditions 
\begin{equation} \label{fq5}
S_{T}(\omega)=\frac{2k_BT^2}{G_{{\rm th}}}\frac{1}{1+ \omega^2 C^2/G_{{\rm th}}^2}.
\end{equation}
At low frequencies we have
\begin{equation} \label{lowfreq}
S_{T}(0)=2k_BT^2/G_{{\rm th}},
\end{equation}
and the spectrum has Lorentzian cut-off at $\omega_c=G_{{\rm th}}/C$. These results hold also for a system coupled to several equilibrium baths, if one takes $G_{\rm th}$ to represent the sum of all the individual thermal conductances to these baths. For the rms fluctuations we obtain the well-known result
$\langle \delta T^2\rangle = \int_{-\infty}^\infty \frac{d\omega}{2\pi} S_T(\omega) = k_B T^2/C$~\cite{lifshitz}.

In a fermionic system, like the electrons in the calorimeter in the present experiment, temperature is coded in the Fermi distribution $f(\epsilon)=[e^{(\epsilon-\mu)/k_{\rm B}T}+1]^{-1}$ which directly determines the readout signal of our thermometer. Here, $\epsilon$ and $\mu$ denote the single particle energy and chemical potential, respectively. We illustrate the calorimeter~\cite{xray, mikko} principle of our experiment and setup in Fig. \ref{fig1}~\cite{brange}. The electron system (absorber), is coupled to the phonon heat bath at constant temperature $T$ via electron-phonon collisions which lead to stochastic exchange of heat, as indicated by the many vertical arrows between the two in Fig.~\ref{fig1}a. The arrows from the left depict the electronic injection of heat under nonequilibrium conditions, fluctuating due to the stochastic nature of tunneling. By attaching a fast thermometer to the absorber, one records its time $t$ dependent temperature fluctuations $\delta T(t)$ as shown by a measured time trace. The actual sample (scanning electron micrograph in Fig.~\ref{fig1}b) is realised as a $L=1~\mu$m long copper normal-metal absorber (brown) connected to three superconducting leads (blue). The right one is a tunnel contact of the thermometer and the other tunnel junction on the left the hot electron injector. The third one pointing down and $50$~nm away from the thermometer, is a direct clean metal-to-metal contact grounded at the sample stage. It provides a fixed chemical potential for the absorber and induces proximity superconductivity to the thermometer facilitating its proper operation. The measuring setup for the thermometer junction shown on the right side of Fig.~\ref{fig1}b consists of a parallel on-chip $LC$ resonator, coupled to input $V_1$ and output $V_2$ rf lines, operating at $f_0=620$ MHz which also admits DC biasing at voltage $V_{\rm th}$. The measured signal $S_{21}$ obtained from the ratio of $V_2/V_1$, yields the conductance of the thermometer junction. It is typically measured at $10$ kHz sampling rate in order to acquire statistics of temporal temperature of the absorber.  
\begin{figure}
	\centering
	\includegraphics [width= 8.5 cm] {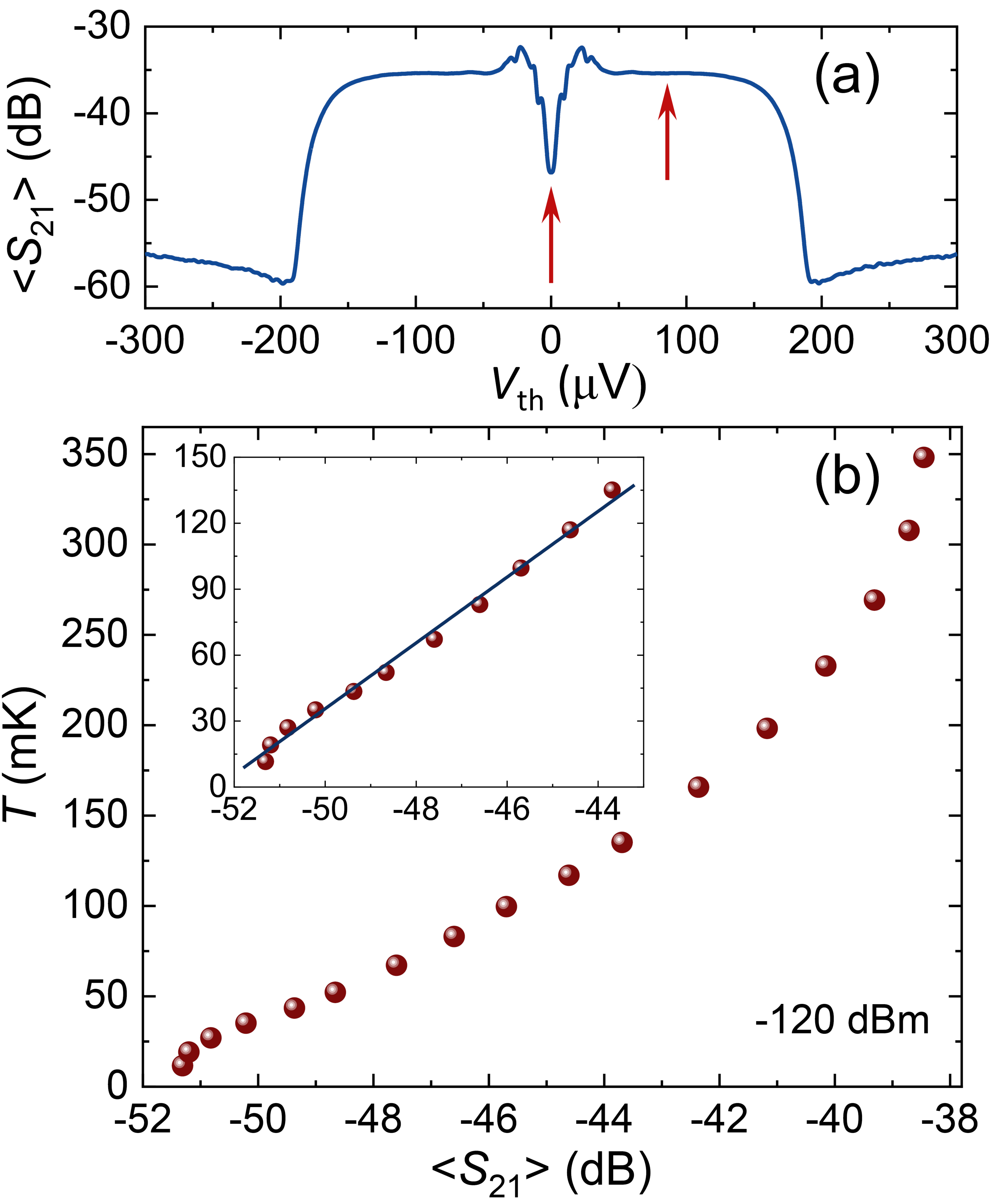}
	\caption{The transmission measurement of the RF thermometer. (a) Wide bias range transmission $\langle S_{21} \rangle$ averaged over 100 repetitions at each bias point $V_{\rm th}$ at bath temperature $T\sim~100$~mK and $-120$~dBm input power. The two red arrows indicate the working points for actual ZBA thermometry at $V_{\rm th}=0$ and background measurement at $V_{\rm th}=85~\mu$V, respectively. (b) The thermometer calibration against the bath temperature $T$ in equilibrium. The inset shows the low temperature end together with the linear fit used for the temperature fluctuation measurements.}
	\label{fig2}
\end{figure}

In order to calibrate the thermometer we measure $S_{21}$ averaged over typically $1$~s time interval at different bath temperatures of the cryostat, traceable to primary Coulomb blockade thermometry CBT. An example of dependence of thus obtained $\langle S_{21} \rangle$ on $V_{\rm th}$ is shown on a wide bias range in Fig.~\ref{fig2}a. The drop of $\langle S_{21} \rangle$ at about $\pm 200~\mu$V is due to the superconducting gap $\Delta$ in aluminum. The main feature, the zero bias anomaly (ZBA) at $V_{\rm th}=0$ which is indicated by the central red arrow, presents the basis of our thermometer. This dip originates from proximity induced supercurrent due to the presence of clean contact. Now it is placed $50$~nm away from the tunnel junction which is to be contrasted to $500$ nm in our earlier work~\cite{bj1}; this way the sensitivity of the thermometer is enhanced substantially. Quantitatively, the temperature dependence of the transmission $\langle S_{21} \rangle$ at this dip is depicted in Fig.~\ref{fig2}b. It manifests approximately linear dependence at sub $200$~mK down to below $20$~mK temperatures, emphasised by the zoom in the inset of this figure. Owing to the competing quasiparticle tunneling, there is eventually back-bending of the characteristics at temperatures above $300$~mK; this leads to temporal loss of sensitivity in this temperature range. Depending on the range of interest, we employ either linear or nonlinear calibration to convert $\langle S_{21} \rangle$ to temperature. As tested and demonstrated in Ref.~\cite{bj1}, the temperature measured by $\langle S_{21} \rangle$ of the ZBA thermometer in a similar setup is that of the absorber electrons that we indeed want to monitor. 
\begin{figure}
	\centering
	\includegraphics [width= 8.5 cm] {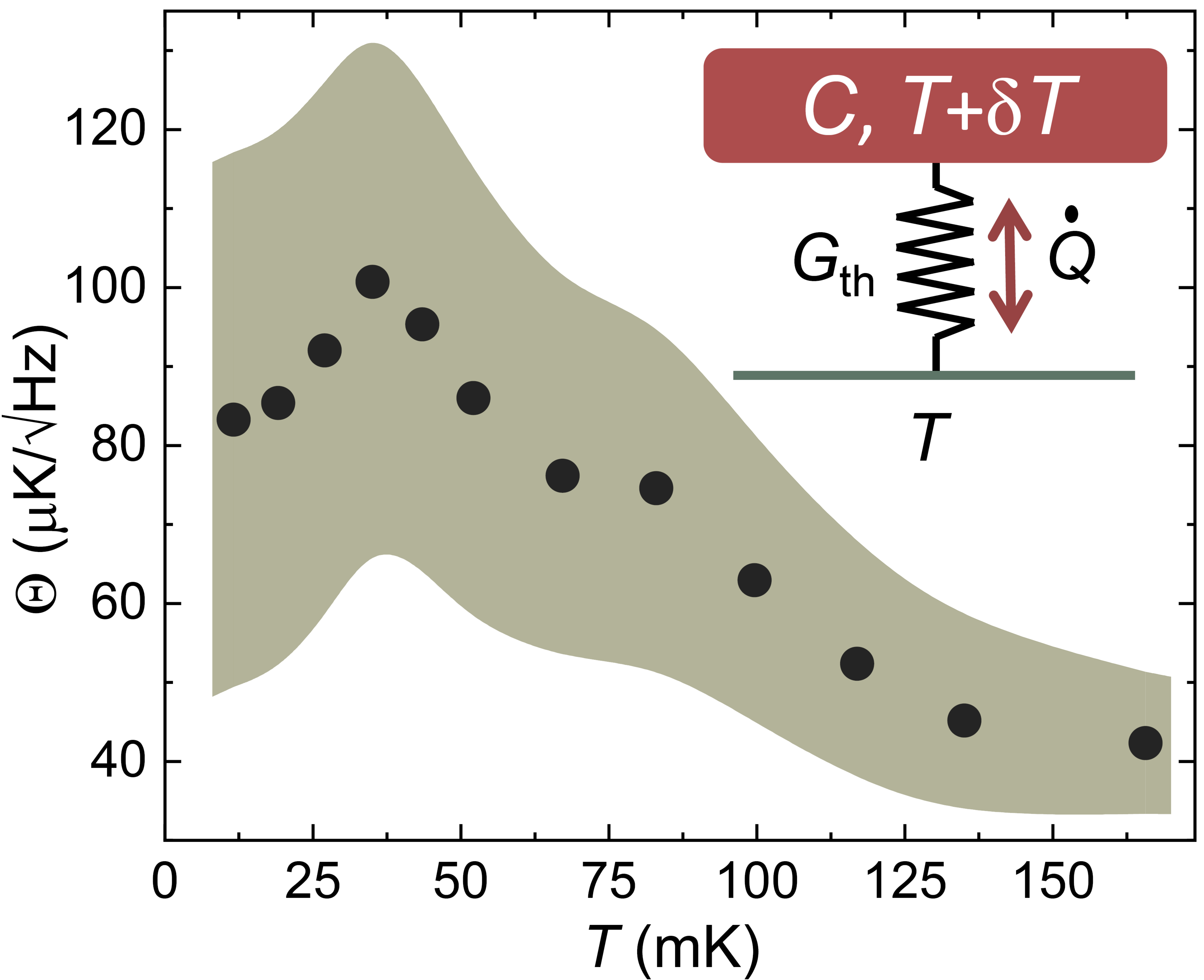}
	\caption{Temperature fluctuations in equilibrium. Measured low frequency fluctuations $\Theta=\sqrt{S_T}$ at different bath temperatures. The symbols are the measured data with the mean instrumental background noise subtracted as described in the Methods section. The shaded area covers the uncertainty in this subtraction process. The inset shows schematically the thermal model including the absorber with finite heat capacity $C$ coupled to the heat bath at temperature $T$ via thermal conductance $G_{\rm th}$. The latter includes various contributions such as electron-phonon coupling and the photonic coupling to the environment. According to the fluctuation dissipation theorem the heat current $\dot{Q}$ fluctuates and this leads to variations $\delta T$ of the absorber temperature.}
	\label{fig3}
\end{figure}
Time domain measurements allow detecting temporal fluctuations of the quantity of interest. In our case we monitor $S_{21}(t)$, yielding the instantaneous temperature of the absorber at $10$~kHz sampling rate over a chosen time interval. We collect data under given conditions typically for up to $1$ hour. As a result we obtain the total fluctuations (variance) $\langle \delta S_{21,{\rm tot}}^2 \rangle$ in a bandwidth of $\Delta f\approx 10$~kHz. This signal is composed of the amplifier and other instrumental noise $\langle \delta S_{21,{\rm bg}}^2 \rangle$ ("bg" stands for background), in addition to the noise of interest from the actual sample, $\langle \delta S_{21}^2 \rangle$=$\langle \delta S_{21,{\rm tot}}^2 \rangle$-$\langle \delta S_{21,{\rm bg}}^2 \rangle$. Here we assume uncorrelated noise from the different sources. The way we determine the $\langle \delta S_{21,{\rm bg}}^2 \rangle$ is explained in the Methods section. Our quantitative results depend critically on the precision of determining this background noise. Taking the linear calibration as in the inset of Fig.~\ref{fig2}b, with the responsivity $\mathbb R\equiv |d\langle S_{21} \rangle /dT|$, we have for the temperature of the absorber $\langle \delta T^2 \rangle=\mathbb R^{-2} \langle \delta S_{21}^2 \rangle$. We exhibit in Fig.~\ref{fig3} the central quantity in the experiment, low frequency temperature fluctuations $\sqrt{S_T}=\sqrt{\langle \delta T^2 \rangle/2 \Delta f}$ as a function of bath temperature in equilibrium. From now on we denote $\Theta \equiv\sqrt{S_T}$ which can also be associated to the noise-equivalent temperature $\mathbb{NET}_0$, where with subscript $0$ we refer to the fundamental temperature fluctuations discussed here. The data symbols correspond to the averaged bare noise, where the best guess of the background has been subtracted. The shaded area depicts the uncertainty in determining $\Theta$ precisely due to this subtraction. Overall, we observe first increase of $\Theta$ upon lowering $T$ and then gradual turn down of it at the lowest temperatures. The dominant contributions to $G_{th}$ arise from electron-phonon coupling at higher temperatures and radiative heat transfer by thermal photons~\cite{schmidt} towards low $T$ as
\begin{equation} \label{Gth}
G_{th}=5\Sigma \mathcal{V} T^4+\alpha g T.
\end{equation}
Here $\Sigma$, $\mathcal{V}$ are electron-phonon coupling constant~\cite{Wellstood} and volume of the absorber, respectively. For the photonic contribution~\cite{schmidt}, $G_Q=gT$ is the quantum of thermal conductance with $g=\pi k_{\rm B}^2/6\hbar$. We assume the coupling coefficient $\alpha$ to have values $\ll 1$. Equation \eqref{lowfreq} predicts then 
\begin{eqnarray}\label{tdep}
\Theta=\sqrt{\frac{2k_{\rm B}}{5\Sigma \mathcal{V}}}T^{-1}~~{(\rm high~}T)\nonumber\\\Theta=\sqrt{\frac{2k_{\rm B}}{\alpha G_Q}}T^{1/2}~~{(\rm low~}T),
\end{eqnarray}
with cross-over between the two regimes at the temperature $T_{\rm co}=(\frac{\alpha g}{10\Sigma \mathcal{V}})^{1/3}$. Using the literature value $\Sigma=2\times 10^9$ $\rm WK^{-5}m^{-3}$~\cite{Libin}, the measured volume $\mathcal{V}=1.0 \times 10^{-21}$ $\rm m^3$ and an educated guess $\alpha \simeq 10^{-3}$~\cite{timofeev} according to earlier investigations, we obtain a predicted $\Theta$ versus $T$. Our simple model above predicts a maximum of $\Theta$ at $\sim~35$~mK with the value of about $30~\mu{\rm K}/\sqrt{\rm Hz}$. This is outside the shaded area of $60-130~\mu{\rm K}/\sqrt{\rm Hz}$ of the measured signal in Fig. \ref{fig3}. A possible origin of this discrepancy lies in that we assume the absorber
to be in the normal state. However, the clean absorber-superconductor
contact leads to a proximity induced superconductivity in the absorber.
This suppresses the density of states around the Fermi level, on the scale
of the Thouless energy $E_{\rm Th}=\hbar D/L^2\sim 10~\mu$eV, resulting in a decreased electron-phonon coupling. Here, $D\sim 0.01$ ${\rm m}^2/{\rm s}$ is the diffusion constant of the Cu film. As a consequence, for electron temperatures $T \lesssim E_{\rm Th}/k_{\rm B} \sim 100$ mK, the thermal conductance $G_\text{th}$ is decreased~\cite{tero2} and, hence, the temperature noise $\Theta$ is increased. 
Furthermore, the temperature calibration, i.e., the responsivity $\mathbb R$ of the thermometer gets more unreliable towards the lowest temperatures. The overall magnitude at $T> T_{\rm co}$, determined by the electron-phonon heat conductance (no fit parameters) is thus in fair agreement with the experiment within the uncertainty of the measurement as explained. The cross-over, more of phenomenological origin, is also consistent with the experiment with the chosen value of $\alpha=10^{-3}$. Another possible contribution to this cross-over may arise from the fact that the fluctuations $\delta T$ of temperature become comparable to $T$ itself in the lowest temperatures.   
\begin{figure}
	\centering
	\includegraphics [width= 8.5 cm] {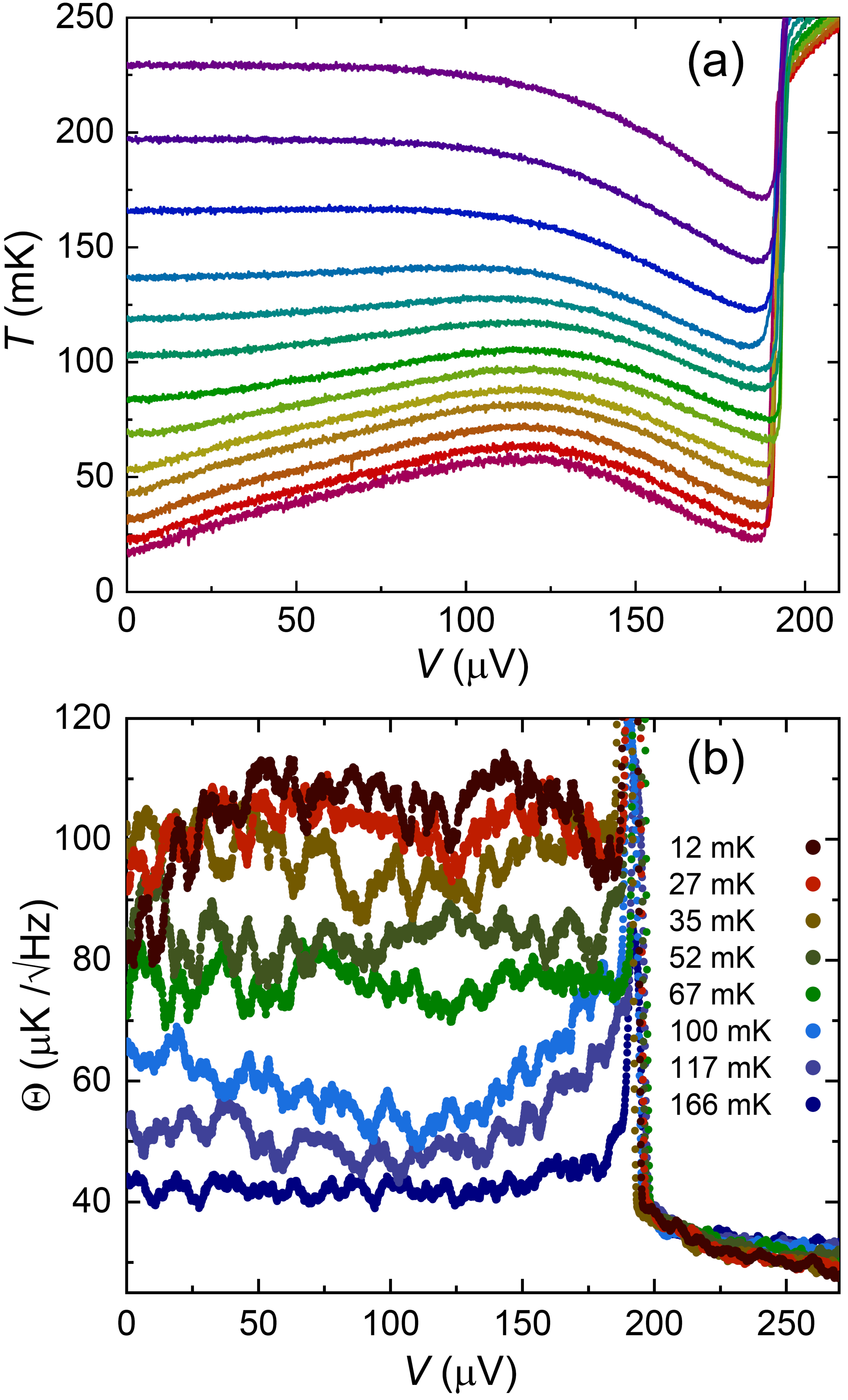}
	\caption{Temperature and its fluctuations under nonequilibrium conditions. (a) Average temperature of the absorber when the injecting junction is biased at different voltages $V$. The data sets correspond to bath temperatures 12, 27, 35, 43, 52, 67, 83, 100, 117, 135, 166, 198, and 233 mK from bottom to top. (b) Nonequilibrium temperature fluctuations at selected temperatures as a function of injector bias.}
	\label{fig4}
\end{figure}

Let us next consider the nonequilibrium fluctuations~\cite{tero,ciliberto,jukka2}. In the measurements up to now the injector junction on the left in Fig.~\ref{fig1}b has been unbiased in order to ensure equilibrium. By applying a voltage $V$ to it, the system can be driven into nonequilibrium. The well-known influence of such biasing of a superconductor-normal metal junction is that it serves as a local refrigerator of the normal-metal absorber thanks to the energy gap of the superconductor, i.e., it acts as an evaporative cooler~\cite{Nahum}. This effect is manifested in the bias dependence of the average temperature of the absorber, obtained from the values of $\langle S_{21} \rangle$ in Fig.~\ref{fig4}a. 

Injecting electrons does not only change the average temperature of the absorber but, due to the stochastic nature of tunneling, it leads to noise of heat current as well~\cite{Dimag}. Quantitatively the low frequency heat current noise is given by 
\begin{eqnarray}\label{shotnoise}
&&S_{\dot{Q}}^{\rm shot}=\frac{1}{e^2R_T}\int dE (E-eV)^2n_{\rm S}(E) \\&& \{ f_{\rm N}(E-eV)[1-f_{\rm S}(E)]+f_{\rm S}(E)[1-f_{\rm N}(E-eV)]\},\nonumber
\end{eqnarray}
where $n_S(E)=|E|/\sqrt{E^2-\Delta^2}\theta(|E|-\Delta)$ denotes the density of states for superconductor and subscripts N and S stand for normal metal and superconductor, respectively. For typical voltages and temperatures in the regime well below the superconducting gap, the injection noise $\sqrt{S_{\dot{Q}}^{\rm in}}$ is exponentially suppressed \cite{brange}. In contrast, the equilibrium noise due to phonons, $\sqrt{S_{\dot{Q}}^{\rm eq}}$, is of a roughly constant magnitude $\sim 10^{-20} {\rm W}/\sqrt{\rm Hz}$. Therefore it is not surprising that the temperature noise in Fig.~\ref{fig4}b does not change
much at sub-gap voltages $V < 200~\mu$V, in particular as
the temperature of the absorber is not changing dramatically in this bias range. For these uncorrelated sources the temperature noise is predicted to obey $S_T=(S_{\dot Q}^{\rm eq}+S_{\dot{Q}}^{\rm in})/{G_{\rm th}}$. The sudden decrease of temperature noise $\Theta$ at $V>200$ $\mu$V is natural since $G_{\rm th}$ increases rapidly when the absorber heats up in this regime (see Fig.~\ref{fig4}a). We consider the sharp peak at the gap (Fig.~\ref{fig4}b) to be an artefact arising from unavoidable voltage noise of injector, which directly transforms to temperature noise due to the strong voltage dependence of temperature at this point.   
\begin{figure} [h!]
	\centering
	\includegraphics [width= 8.5 cm] {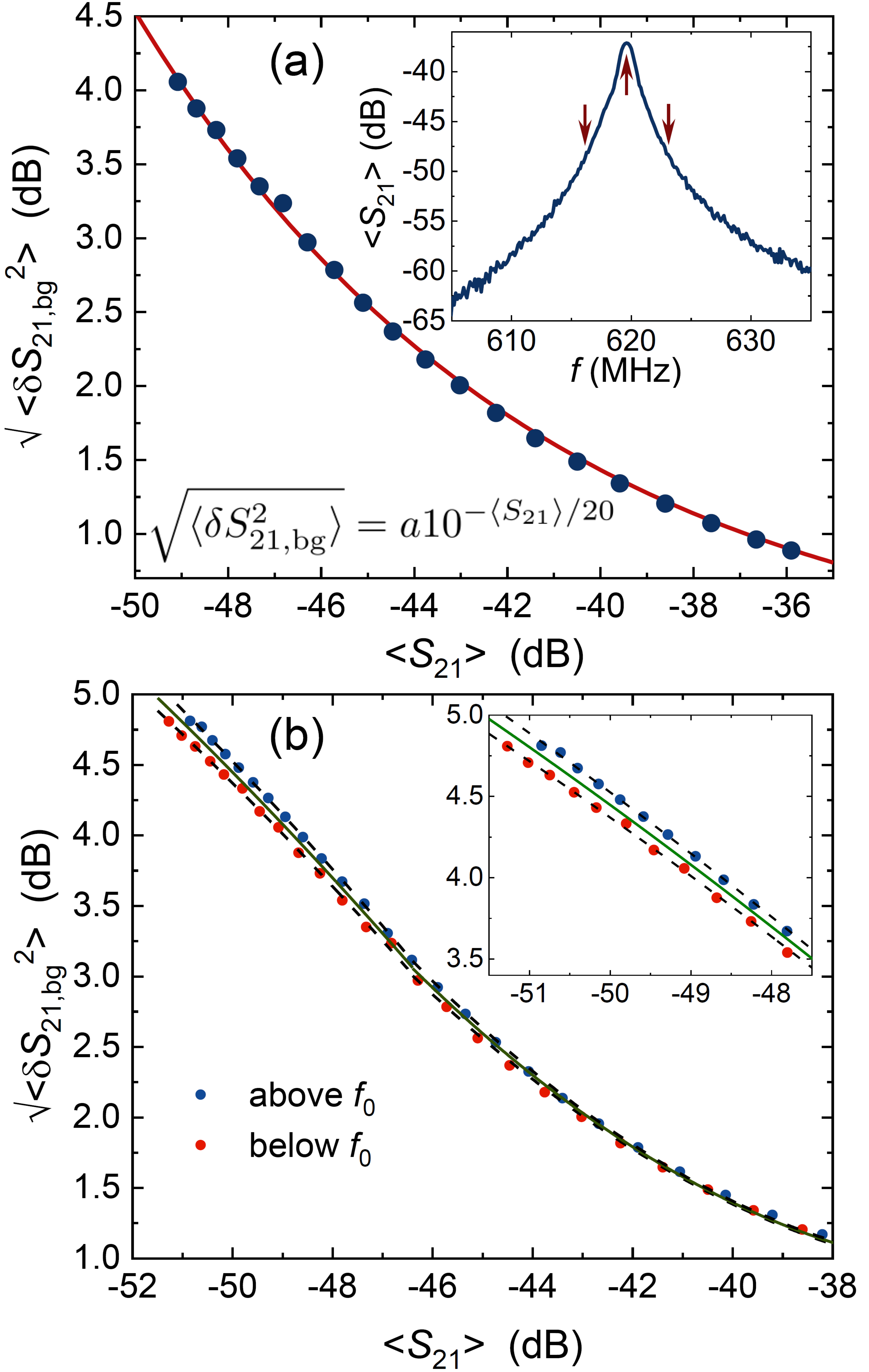}
	\caption{Background noise measurements. All the data are taken outside the zero bias regime of the thermometer and at nonresonant frequencies to exclude the actual noise from the sample. The inset of (a) shows an example of $ \langle S_{21} \rangle$ measured around the resonance frequency indicated by the central upward arrow. The data points in the main frame of (a) depict parametric plot $\sqrt{\langle \delta S_{21,{\rm bg}}^2 \rangle}$ versus $\langle  S_{21} \rangle$ at the bias voltages $V_{\rm th}=85~\mu$V and at frequencies below the resonance down to 614 MHz indicated by a downward arrow. The red solid line shows the predicted dependence yielding the noise temperature of the amplifier of $T_{\rm n}=4.9$~K as the only fit parameter of the curve (constant noise voltage at the input). (b) The full range measurement of the background as in (a) but now both above and below the resonance. The polynomial fits for the two backgrounds separately (black dashed lines) and the average of them (green solid line) are shown, and they define, the mean and the shaded area in Fig.~\ref{fig3}. The inset of (b) is simply the zoom-out of the high attenuation range of the main frame.}
	\label{fig5}
\end{figure}

Finally, {\it what is the temperature that fluctuates?} In fact, it is the parameter of the distribution of the electrons in the absorber that we monitor. It qualifies as temperature for the following reasons. (i) Number of particles is large, about $10^8$. (ii) Due to  fast electron-electron internal relaxation over a time scale of $\sim 10^{-9}$~s~\cite{pothier}, the carriers form a local Fermi-Dirac distribution: all other relaxation rates, most notably the electron-phonon time ($\sim 10^{-5}$~s) are much slower~\cite{viisanen}. Furthermore, the temperature of the absorber is spatially uniform, since the heat diffusion time of electrons in the absorber, $\tau_{\rm diff}=\gamma \rho l^2/\mathcal{L}_0 \sim 10^{-10}$~s is very short. Here, $c=\gamma T$, is the specific heat due to conductance electrons with $\gamma \sim 10^2$ ${\rm Wm^{-3}K^{-2}}$, $\rho\sim 10^{-8}$~$\Omega$m is the resistivity of the Cu, $l=1~\mu$m is the length of the absorber, and $\mathcal{L}_0=2.44\times 10^{-8}~{\rm W\Omega K^{-2}}$ is the Lorenz number.
\section*{Methods}   
\subsection*{Background measurements} 
We measure the instrumental noise dominated by that of the low temperature Caltech CITLF2 cryogenic SiGe low noise amplifier $\langle \delta S_{21,{\rm bg}}^2 \rangle$ by carefully off-tuning the interesting fluctuations from the sample itself. This is achieved by simultaneously (i) biasing the thermometer junction away from the ZBA regime ($V_{\rm th} \simeq 85~\mu$V), and (ii) measuring at frequencies either below or above the resonance at $f_0$. An example of the corresponding parametric background noise measurement, in form $\sqrt{\langle \delta S_{21,{\rm bg}}^2 \rangle}$ versus $\langle S_{21}\rangle$ is presented in Fig. \ref{fig5}. We see a typical increase of noise when the attenuation increases towards left. This dependence can be understood quantitatively by assuming constant voltage noise independent of $\langle S_{21}\rangle$. The measured transmission can be written as 
\begin{equation}\label{s211}
S_{21}=20 \lg ({v}/{\widetilde{v}}),
\end{equation}
where $v$ is the output of the last stage amplifier, $\widetilde{v}=\sqrt{50~{\rm \Omega}\times 1~{\rm mW}}\simeq 224$~mV. Noise of $v$ translates then into variations of $S_{21}$ in linear regime as 
\begin{equation}\label{s212}
\delta S_{21}=\frac{20}{\ln 10} \frac{\delta v}{v},
\end{equation}
and can be written with the help of Eq.~\eqref{s211} for the rms values as
\begin{equation}\label{s213}
\sqrt{\langle \delta S_{21,{\rm bg}}^2 \rangle}=\frac{20}{\ln 10} \frac{\sqrt{\langle \delta v^2 \rangle}}{ \widetilde{v}}10^{-\langle S_{21} \rangle/20}.
\end{equation}
Based on the fit parameter $a$ in Fig. \ref{fig5}a and the total gain of $60$~dB of the amplifier chain, we find the input voltage noise to be $\sim~12$~nV corresponding to the noise temperature of the amplifier of $T_{\rm n}\sim~5$~K which is in line with its specifications by the manufacturer. 

Figure~\ref{fig5}b presents background measurements at frequencies both below and above the resonance over a wide range of attenuation $\langle S_{21} \rangle$. We observe two features that we need to consider when making an accurate evaluation of the $\langle \delta S_{21,{\rm bg}}^2 \rangle$. First, at large attenuations, due to the fact that the changes are not fully linear in the sense of Eq.~\eqref{s212}, the exponential dependence of Eq.~\eqref{s213} is not obeyed strictly. Therefore we resort to polynomial fits in two regimes, to capture the dependence over the full range. Second, there is a weak dependence of the amplifier noise on frequency; thus the data taken below and above the resonance differ from each other slightly. What we do then, e.g., in Fig. \ref{fig3}, is that we take the mean between the two background measurements as the reference and indicate by the shaded area the uncertainty incurred due to the difference between the two extremes. We thus assume that the frequency dependence of the noise is more or less smooth in the narrow range of $\sim~10$~MHz around $f_0$, and interpolate the data accordingly.    

\subsection*{Experimental details}
The sample (Fig.~\ref{fig1}b) was fabricated on standard oxidized Si substrate using Ge process for achieving robust deposition mask~\cite{meschke,pashkin}. The electron-beam lithography was used to pattern the structure for three-angle shadow evaporation of metals. First we deposit $20$~nm of Al making the leads followed by oxidation in pure $O_2$ (1~min at 1~mbar). Next another Al layer of $20$~nm thickness again provides the clean superconducting contact at the distance of $50$~nm from the thermometer junction, and finally we deposit $35$~nm Cu to form the absorber. The resonator is a spiral on a separate chip made of $100$~nm thick Al by simple one angle evaporation. The heart of the measuring setup is shown in Fig.~\ref{fig1}b with inductance $L=100$~nH, $C_1=10.3$~fF and $C_2=59.3$~fF as coupling capacitors, and $C=0.2$~pF. The rest of the RF circuitry follows closely to what is presented in Ref.~\cite{viisanen}.
\section*{Author contributions} 
The experiment was proposed by J. P. and its realization was conceived by all the authors. B. K. performed the experiment, and designed and fabricated the samples. Data analysis and modeling were performed by B. K. and J. P., with contributions on the noise analysis by F. B. and P. S. The manuscript was written by B. K. and J. P.

\section*{acknowledgments}
We acknowledge J. T. Peltonen, E. T. Mannila, O-P. Saira and S. Gasparinetti for technical support, W. Belzig, D. Nikolic, I. Khaymovich, and T. Tuukkanen for discussions and tests of thermometry, and M. Campisi and K. Saito for useful discussions. This work was funded through Academy of Finland grants 297240, 312057 and 303677 and from the European Union's Horizon 2020 research and innovation programme under the European Research Council (ERC) programme and Marie Sklodowska-Curie actions (grant agreements 742559 and 766025). F.B and P.S. were supported by the Swedish VR. We acknowledge the facilities and technical support of Otaniemi research infrastructure for Micro and Nanotechnologies (OtaNano).


\begin{thebibliography}{99}
\bibitem{johnson} Johnson, J. B. Thermal agitation of electricity in conductors. Phys. Rev. {\bf 32}, 97-109 (1928).
 
\bibitem{nyquist} Nyquist, H. Thermal agitation of electric charge in conductors. Phys. Rev. {\bf 32}, 110-113 (1928).

\bibitem{lifshitz} Lifshitz, E. M. \& Pitaevskii, L. P. Statistical physics, Part I (Pergamon, Oxford, 1980).

\bibitem{kittel} Kittel, C. Temperature fluctuation: an oxymoron. Phys. Today {\bf 41}, 93 (1988).

\bibitem{mandelbrot} Mandelbrot, B. B. Temperature Fluctuation: a well‐defined and unavoidable notion. Phys. Today {\bf 42}, 71-73 (1989).

\bibitem{lipa} Chui, T. C. P., Swanson, D. R., Adriaans, M. J., Nissen, J. A. \& Lipa, J. A. Temperature Fluctuations in the canonical ensemble. Phys. Rev. Lett. {\bf 69}, 3005-3008 (1992).

\bibitem{lipa2} Day, P., Hahn, I., Chui, T. C. P., Harter, A. W., Rowe, D. \& Lipa, J. A. The fluctuation-imposed limit for temperature measurement. J. Low Temp. Phys. {\bf 107}, 359-370 (1997).

\bibitem{xray} Stahle, C. K., McCammon, D. \& Irwin, K. D. Quantum Calorimetry. Phys. Today {\bf 52}, 8, 32 (1999).

\bibitem{mikko} Govenius, J., Lake, R. E., Tan, K. Y. \& M\"ott\"onen, M. Detection of Zeptojoule Microwave Pulses Using Electrothermal Feedback in Proximity-Induced Josephson Junctions. Phys. Rev. Lett. {\bf 117}, 030802 (2016).


\bibitem{brange} Brange, F., Samuelsson, P., Karimi, B. \& Pekola, J. P. Nanoscale quantum calorimetry with electronic temperature fluctuations. Phys. Rev. B {\bf 98}, 205414 (2018).

\bibitem{bj1} Karimi, B. \& Pekola, J. P. Noninvasive thermometer based on the zero-bias anomaly of a superconducting junction for ultrasensitive calorimetry. Phys. Rev. Appl. {\bf 10}, 054048 (2018).

\bibitem{Libin} Wang, L. B., Saira, O.-P., Golubev, D. S. \& Pekola, J. P. Crossover between electron-phonon and boundary resistance limited thermal relaxation in copper films, to be published.

\bibitem{timofeev} Timofeev, A. V., Helle, M., Meschke, M., M\"ott\"onen, M. \& Pekola, J. P. Electronic refrigeration at the quantum limit. Phys. Rev. Lett. {\bf 102}, 200801 (2009).


\bibitem{brange2} van den Berg, T. L., Brange, F., \& Samuelsson, P. Energy and temperature fluctuations in the single electron box. New J. Phys. {\bf 17}, 075012 (2015).


\bibitem{Wellstood} Wellstood, F. C., Urbina, C., \& Clarke, J. Hot-electron effects in metals. Phys. Rev. B {\bf 49}, 5942-5955 (1994).

\bibitem{schmidt} Schmidt, D. R., Schoelkopf, R. J., \& Cleland, A. N. Photon-mediated thermal relaxation of electrons in nanostructures. Phys. Rev. Lett. {\bf 93}, 045901 (2004).

\bibitem{tero2} Heikkil\"a, T. T., \& Giazotto, F. Phase sensitive electron-phonon coupling in a superconducting proximity structure. Phys. Rev. B {\bf 79}, 094514 (2009).

\bibitem{Nahum} Nahum, M., Eiles, T. M., \& Martinis, J. M. Electronic microrefrigerator based on a
normal-insulator-superconductor tunnel junction. Appl. Phys. Lett. {\bf 65}, 3123 (1994).

\bibitem{jukka2} Averin, D. V. \& Pekola, J. P. Violation of the fluctuation-dissipation theorem in time-dependent mesoscopic heat transport. Phys. Rev. Lett. {\bf 104}, 220601 (2010).

\bibitem{tero} Heikkil\"a, T. T. \& Nazarov, Y. V. Statistics of temperature fluctuations in an electron system out of equilibrium. Phys. Rev. Lett. {\bf 102}, 130605 (2009).

\bibitem{ciliberto} Ciliberto, S., Imparato, A., Naert, A. \& Tanase, M. Heat flux and entropy produced by thermal fluctuations. Phys. Rev. Lett. {\bf 110}, 180601 (2013).

\bibitem{Dimag} Golubev, D. \& Kuzmin, L. Nonequilibrium theory of a hot-electron bolometer with normal metal-insulator-superconductor tunnel junction. J. Appl. Phys. {\bf 6464}, 6464 (2001).

\bibitem{pothier} Pothier, H., Gueron, S., Birge, N. O., Esteve, D. \& Devoret, M. H. Energy distribution function of quasiparticles in mesoscopic wires. Phys. Rev. Lett. {\bf 79}, 3490-3493 (1997).

\bibitem{viisanen} Viisanen, K. L. \& Pekola, J. P. Anomalous electronic heat capacity of copper nanowires at sub-Kelvin temperatures. Phys. Rev. B {\bf 97}, 115422 (2018).


\bibitem{meschke} Meschke, M., Kemppinen, A., \& Pekola, J. P. Accurate Coulomb blockade thermometry up to 60 kelvin. Phil. Trans. R. Soc. A {\bf 374}, 20150052 (2015).


\bibitem{pashkin} Pashkin, Y., Nakamura, Y., \& Tsai, J.-S. Implementation of single-electron transistor with resistive gate. Jpn. J. Appl. Phys. {\bf 38}, 406-409 (1999).


\end{thebibliography}
\end{document}